\documentclass[runningheads,a4paper,11pt]{llncs}

\usepackage[top=2.5cm,bottom=2.3cm,hmargin=2.5cm]{geometry}

\usepackage{amssymb,graphicx,url}
\usepackage[english]{babel}
\usepackage{multirow}
\usepackage[dvipsnames,svgnames]{xcolor}
\usepackage[ruled,vlined,linesnumbered]{algorithm2e}
\usepackage{xspace}
\usepackage{pgfplots}
\usepackage{pgfplotstable}

\title{High-dimensional approximate nearest neighbor: \\
\kd Generalized Randomized Forests}
\titlerunning{ANN in High Dimensions: kd-GeRaF} 

\author{Yannis Avrithis\thanks{The first two authors are partially supported
by the European Social Fund and Greek National Fund through
the National Strategic Reference Framework, research funding program
``ARISTEIA'', project ``ESPRESSO".}%
\and Ioannis Z.~Emiris$^{\star}$\and Georgios Samaras}
\authorrunning{Y.~Avrithis, I.Z.~Emiris, and G.~Samaras}

\urldef{\mailsa}\path|iavr@image.ntua.gr,emiris@di.uoa.gr,georgesamarasdit@gmail.com|
\newcommand{\keywords}[1]{\par\addvspace\baselineskip
	\noindent\keywordname\enspace\ignorespaces#1}

\institute{Dept of Informatics \& Telecommunications, University of Athens, Greece\\
\mailsa}

\newcommand{\kd}{\emph{k}-d\space}
\newcommand{\kgeraf}{\kd GeRaF}
\newcommand{\geraf}{{\tt GeRaF}\xspace}
\newcommand{\att}[1]{{#1}}

\newcommand{\head}[1]{{\medskip\noindent\emph{#1}}}

\newtheorem{defn}{Definition}

\def\real{{\mathbb R}}

\makeatletter
\DeclareRobustCommand\onedot{\futurelet\@let@token\@onedot}
\def\@onedot{\ifx\@let@token.\else.\null\fi\xspace}
\def\eg{\emph{e.g}\onedot} 
\def\ie{\emph{i.e}\onedot}

\makeatother

\begin{document}

\mainmatter
\maketitle

\begin{abstract}
We propose a new data-structure, the generalized randomized \kd forest, or \kgeraf, for approximate nearest neighbor searching in high dimensions. In particular, we introduce new randomization techniques to specify a set of independently constructed trees where search is performed simultaneously, hence increasing accuracy. We omit backtracking, and we optimize distance computations, thus accelerating queries. We release public domain software \geraf and we compare it to existing implementations of state-of-the-art methods including BBD-trees, Locality Sensitive Hashing, randomized \kd forests, and product quantization. Experimental results indicate that our method would be the method of choice in dimensions \att{around~1,000, and probably up to~10,000}, and pointsets of cardinality up to a few hundred thousands or even one million; this range of inputs is encountered in many critical applications today. For instance, we handle a real dataset of $10^6$ images represented in 960 dimensions with a query time of less than $1$sec on average and 90\% responses being true nearest neighbors.
\keywords{data-structure, randomized tree, space partition, geometric search, open software, practical complexity}
\end{abstract}
\section{Introduction}
\label{sec:intro}

After a couple of decades of work, Nearest Neighbor Search remains a fundamental optimization problem with both theoretical and practical open issues today, in particular for large datasets in dimension well above~100. An exact solution using close to linear space and sublinear query time is impossible, hence the importance of approximate search, abbreviated {\em NNS}. We focus on the Euclidean metric but extensions to other metrics should be possible.

\begin{defn}
\label{Danns}
Given a finite dataset $X\subset\real^d$ and real $\epsilon > 0$, $x^* \in X$ is an $\epsilon$-approximate nearest neighbor of query $q \in \real^d$, if $dist(q, x^*) \le (1 + \epsilon) dist(q, x)$ for all $x\in X$. For $\epsilon = 0$, this reduces to exact NNS.
\end{defn}

Despite a number of sophisticated methods available, it is still open which is best for various ranges of the input parameters. Here, we propose a practical data-structure, generalizing \kd trees, which should be the method of choice in dimension roughly in the range of \att{1,000 to~10,000 and inputs of a few hundred thousand points, and up to a million}. By taking advantage of randomization and new algorithmic ideas, we offer a very competitive open software for (approximate) NNS in this range of inputs, which provides a good trade-off between accuracy and speed. Our work also sheds light into the efficiency of \kd trees, which is one of the most common data structures but whose complexity analysis is far from tight.

High-dimensional NNS arises naturally when complex objects are represented by vectors of $d$ scalar features. NNS tends to be one of the most computationally expensive parts of many algorithms in a variety of applications, including
computer vision
, knowledge discovery and data mining
, pattern recognition and classification
, machine learning
, data compression
, multimedia databases
, document retrieval
~and statistics
~\cite{JeDS11,MuLo14,PCISZ07,WSSJ14,WeTF08}. Large scale problems are quite common in such areas, for instance more than $10^7$ points and more than $10^5$ dimensions~\cite{PAHS12}.


\head{Previous work.}
There are many efficient approaches to NNS. We focus on the most competitive ones, with emphasis on practical performance.

An important class of methods consists in data-dependent methods, where the decisions taken for space partitioning are based on the given data points.
The Balanced Box Decomposition (BBD) tree~\cite{AMN+98} is a variant of the quadtree,
closely related to the fair-split tree.
It has $O(\log{n})$ height, and subdivides space into axis-aligned hyper-rectangles,
containing one or more points with bounded aspect ratio.
It achieves query time $O(d^{d+1} \log n / \epsilon^d)$,
using space in $O(dn)$, and preprocessing time in $O(d n \log n)$.
The implementation in library {\tt ANN}\footnote{\url{http://cs.umd.edu/~mount/ANN}} seems to be the most competitive method for roughly $d<100$. Recently, a novel dimensionality reduction method has been combined with BBD-trees to yield NNS with optimal space requirements and sublinear query time \cite{AnEmPs15socg}.

The performance of BBD-trees in practice is comparable to that of \kd trees. The latter lack a tight analysis but it is known that search becomes almost linear in $n$ for large $d$ because of backtracking. Randomization is a powerful idea: in~\cite{SAnHar08}, a random isometry is used with \kd trees; in~\cite{Vemp12}, tree height is analyzed under random rotations; Random Projection trees~\cite{DasFre08} take another tack. R-trees and their variants are most frequent in database applications: they are comparable in performance to \kd trees, but lack complexity and error bounds.

\kd trees are probably the most common data-structure for NNS, having implementations in libraries {\tt ANN}, with performance comparable to BBD-trees, and {\tt CGAL}, which is competitive only for small inputs. A successful contribution has been library {\tt FLANN}~\cite{MujaLo09,MuLo14}, considered state-of-the-art for $d$ about~100; the method has been most successful on SIFT image descriptors with $d=128$. {\tt FLANN}\footnote{\url{http://cs.ubc.ca/research/flann}} constructs a forest of up to~6 randomized \kd trees and performs simultaneous search in all trees. It chooses the split coordinates adaptively but all leaves contain a single point. The implementation adopts some optimization techniques, such as unrolling the loop of distance computation, but our software goes significantly further in this direction.

In high dimensional space, tree-based data structures are affected by the curse of dimensionality, i.e., either the running time or the space requirement grows exponentially in $d$. An important method conceived for high dimensional data is Locality Sensitive Hashing (LSH). LSH induces a data independent space partition and is dynamic, since it supports insertions and deletions. The basic idea of LSH is to hash the points of the data set so as to ensure that the probability of collision is much higher for objects that are close to each other than for those that are far apart.
The existence of such hash functions depends on the metric space. In general, LSH requires roughly $O(dn^{1+\rho})$ space and $O(dn^{\rho})$ query time for some $\rho \in (0,1)$. It is known~\cite{AI08} that in the Euclidean case, it is possible to bound $\rho$ by $\rho\leq{1}/{(1+\epsilon)^2}$.
One implementation that we use for comparisons is in library {\tt E$^2$LSH}\footnote{\url{http://www.mit.edu/~andoni/LSH}}.


A different hashing approach is to represent points by short binary codes to approximate and accelerate distance computations. Recent research on learning such codes from data distributions is very active~\cite{WSSJ14}. A more general approach is to use any discrete representation of points, again learned from data points. A popular approach is product quantization (PQ)~\cite{JeDS11}, which both compresses data points and provides for fast asymmetric distance computations, where points remain compressed but queries are not. A powerful non-exhaustive search method inspired by PQ is the inverted multi-index~\cite{BaLe12}. There are several recent extensions, and the current state of the art in up to $10^9$ points in 128 dimensions is locally optimized product quantization (LOPQ)~\cite{KaAv14}.


\head{Contribution.} Our main contribution is to propose a new, randomized data-structure for NNS, namely the \kd \emph{Generalized Randomized Forest} (\kgeraf), which generalizes the \kd tree in order to perform fast and accurate NNS in high dimensions and dataset cardinality in thousands or millions. We employ adaptive and randomized algorithms for choosing the split coordinate, and further randomization techniques to build a number of independent \kd trees. We also provide automatic configuration of the parameters governing tree construction and search. All trees are searched simultaneously, with no need for backtracking.
We examine alternative ideas, such as random shuffling of the points, random isometries, leaves with several points,
and methods for accelerating distance computation. By keeping track of encountered points, we avoid repeated computations~\cite{MuLo14}.

We have implemented all of the above techniques within a public domain C++ software, \geraf. This has also allowed us to experiment with different alternatives and provide a simple yet effective automatic parameter configuration. We compare to the main existing alternative libraries on a number of synthetic and real datasets of varying dimensionality and cardinality. We have experimented with parameters of all methods and observed the difficulty, in general, to optimize them. Automatic configuration works fine for \geraf, which is the fastest in building. \geraf also scales very well, even for $d=10^4$ or $n=10^6$, and, at the same accuracy, it is faster than competition for $d$ roughly in the range $(10^3 , 10^4)$, and $n$ in the hundreds of thousands or millions.

The paper is structured as follows. Section~\ref{sec:method} discusses the data structure and method, including randomization factors, building the forest, searching, and improvements that we introduce. Section~\ref{sec:impl} focuses on more technical implementation issues. Section~\ref{sec:exp} presents experimental evaluation and comparisons, while conclusions are drawn in Section~\ref{sec:discuss}.
\section{The \kgeraf}
\label{sec:method}

The limitations of a single \kd tree for high $d$ are overcome by searching multiple, randomized trees, simultaneously. This section discusses randomization, and algorithms for parameter configuration, building, and searching. Overall, $m$ different randomized \kd trees are built, each with a different structure such that search in the different trees is independent; \ie, neighboring points that are split by a hyperplane in one, are not split in another. Search is simultaneous in the $m$ trees, \ie, nodes from all trees are visited in an order determined by a shared priority queue. There is no backtracking, and search terminates when $c$ leaves are visited.


\subsection{Randomization}
\label{sec:random}

The key insight is to construct substantially different trees, by randomization. Multiple independent searches are subsequently performed, increasing the probability of finding approximate nearest neighbors. Randomization amounts to either generating a different randomly transformed pointset per tree (\eg, rotation or shuffling), or choosing splits at random at each node (\eg, split dimension or value). As discussed below, we investigate four randomization factors, which we use either independently or in combination.



\head{Rotation.} For each \kd tree, we randomly rotate the input pointset or, more generally, apply a different isometry~\cite{SAnHar08}. Each resulting tree is thus based on a different set of dimensions. Only the transformation matrix $R$ is stored for each tree, and not the rotated set. In fact, not even the entire matrix needs to be stored, as discussed in section~\ref{sec:build}. During search, the query is rotated using $R$ before descending each tree. However, distances are computed between the original stored points and the original query.

\head{Split dimension.} In a conventional \kd tree, the pointset is halved at each node along one dimension; dimensions are examined in order even for high $d$. Here, we find the $t$ dimensions of highest variance for the input set and then choose uniformly at random one of these $t$ dimensions at each node. Thus, different trees are built from the given pointset.

\head{Split value.} The default split value in a conventional \kd tree is the median of the coordinates in the selected split dimension. {\tt FLANN} uses the mean for reasons of speed. Here, we compute the median, which would yield a perfect tree, and then randomly perturb it~\cite{PCISZ07}. In particular, the split value equals the median plus a quantity $\delta$ uniformly distributed in $[-{3\Delta}/{\sqrt{d}}, {3\Delta}/{\sqrt{d}}]$, where $\Delta$ is the diameter of the current pointset; $\delta$ is computed at every node during building~\cite{Vemp12}.

\head{Shuffling.} When computing the split value at each node in a conventional \kd tree, the current pointset at the node is used.
Even if the split value is randomized, it is still possible that the same point is chosen if the same coordinate value occurs more than once in the selected dimension. This is particularly common when points are quantized; for instance, SIFT vectors are typically represented by one byte per element. We thus randomly shuffle points at each tree. Hence, different splits occur despite ties.


\subsection{Building}
\label{sec:build}

\begin{algorithm}[t]
\SetFuncSty{textsc}
\SetDataSty{emph}
\newcommand{\expr}[1]{$\langle${#1}$\rangle$}
\newcommand{\commentsty}[1]{{\color{DarkGreen}#1}}
\SetCommentSty{commentsty}
\SetKwComment{Comment}{$\triangleright$ }{}
\SetKwBlock{Block}{function}{}
\SetKwInOut{Input}{input}
\SetKwInOut{Output}{output}
\SetKw{Return}{return}
\SetKwData{Node}{node}
\SetKwData{Leaf}{leaf}
\SetKwFunction{Build}{build}
\SetKwFunction{Variance}{variance}
\SetKwFunction{Median}{median}
\SetKwFunction{Split}{split}

\Input
{pointset $X$, \#trees $m$, \#split-dimensions $t$, max \#points per leaf $p$
}
\Output{randomized \kd forest $F$}
\Begin
{
	$V \gets$ \expr{\Variance of $X$ in every dimension} \\
	$D \gets$ \expr{$t$ dimensions of maximum variance $V$} \\
	$F \gets \emptyset$  \Comment*[f]{forest} \\
	\For{$i \gets 1$ \KwTo $m$}
	{
		$f \gets$ \expr{random transformation}  \Comment*[f]{isometry, shuffling} \\
		$F \gets F \cup (f,\Build(f(X))$) \Comment*[f]{build on transformed X, store $f$}
	}
	\Return $F$
}

\BlankLine

\Block({$\Build(X)$} \Comment*[f]{recursively build tree (node/leaf)})
{
	\If(\Comment*[f]{termination reached}){$|X| \le p$}{\Return $\Leaf(X)$}
	\Else(\Comment*[f]{split points and recurse})
	{
		$s \gets$ \expr{one of dimensions $D$ at random} \\
		$v \gets $ \expr{\Median of $X$ in dimension $s$} \\
		$(L,R) \gets$ \expr{\Split of $X$ in dimension $s$ at value $v$} \\
		\Return $\Node(c,v,\Build(L),\Build(R))$ \Comment*[f]{build children on $L,R$}\\
	}
}
\caption{\kgeraf: building\label{alg:build}}
\end{algorithm}

The overall building algorithm for \kgeraf, consisting of $m$ trees, is outlined in Alg.~\ref{alg:build} (Appendix). For simplicity, only the random split dimensions are included, while the split value is the standard median. There is a random data transformation $f$ per tree, which may include either an isometry, shuffling, or both; in case of an isometry, it is stored for use during search.

Given a dataset $X$, the $t$ dimensions of maximum variance, say $D$, are computed. For each tree, $X$ is transformed according to a different function $f$ and then the tree is built recursively. At each node, one dimension (coordinate), say $s$, is chosen uniformly at random from $D$ and $X$ is split at the median in $s$. The two subsets of $X$, say $L, R$, are then recursively given as input datasets to the two children of the node. The split node so constructed contains the split dimension $s$ and the split value $v$. Splitting terminates when fewer than $p$ points are found in the dataset, in which case the point indices are just stored in a leaf node. When $n$ is much higher than $d$, the bottleneck of the algorithm is finding the median, which is $O(n)$ on average. Otherwise, the bottleneck is computing the variance per dimension, which is $O(d)$. The space requirement for the entire data structure is $O(nd)$ for the data points and $O(nm)$ for the trees, including both nodes and indices to points, for a total of $O(n(d+m))$.

Each random isometry can be a rotation \cite{Vemp12} or reflection, and in general requires the generation of a random orthogonal matrix $R$. We rather use an elementary Householder reflector $P$ for efficiency~\cite{SAnHar08}. In particular, given unit vector $u\in\real^d$ normal to hyperplane $H$, the orthogonal projection of a point $x$ onto $H$ is $x - (u^\top x) u$. Its reflection across $H$ is twice as far from $x$ in the same direction, that is, $y = x - 2(u^\top x) u = Px$, where $P = I - 2 u u^\top.$ Although $P$ is orthogonal, the computation of reflection $Px$ is $O(n)$, involving a dot product and an element-wise multiplication and addition. This is because $u u^\top$ is of rank one. We only need to store vector $u$ for each tree.

\begin{algorithm}[t]
\SetFuncSty{textsc}
\SetDataSty{emph}
\newcommand{\expr}[1]{$\langle${#1}$\rangle$}
\newcommand{\commentsty}[1]{{\color{DarkGreen}#1}}
\SetCommentSty{commentsty}
\SetKwComment{Comment}{$\triangleright$ }{}

\SetKwBlock{Block}{function}{}
\SetKwInOut{Input}{input}
\SetKwInOut{Output}{output}
\SetKw{Return}{return}
\SetKw{Break}{break}

\SetKwFunction{True}{true}
\SetKwFunction{False}{false}

\SetKwData{Check}{check}
\SetKwData{Left}{left}
\SetKwData{Right}{right}
\SetKwData{Node}{node}
\SetKwData{Leaf}{leaf}

\SetKwFunction{Descend}{descend}
\SetKwFunction{Max}{max}
\SetKwFunction{Points}{points}
\SetKwFunction{Dist}{dist}

\SetKwFunction{Init}{init}
\SetKwFunction{Empty}{empty}
\SetKwFunction{Insert}{insert}
\SetKwFunction{Extract}{extract-min}

\Input{query point $q$, forest $F$, \#neighbors $k$, max \#leaf-checks $c$}
\Output{$k$ nearest points}
\Begin
{
	$Q.\Init()$  \Comment*[f]{min-priority queue, initially empty} \\
	\For{$i \gets 1$ \KwTo $m$}
	{
		$\Descend(q,F[i],\False)$  \Comment*[f]{descend $i$-th tree, store path in $Q$, no checks} \\
	}
	$\ell \gets 0$  \Comment*[f]{\# of leaves checked} \\
	$H.\Init(k)$  \Comment*[f]{min-heap of size $k$} \\
	\While{$\neg Q.\Empty() \wedge \ell < c/(1+\epsilon)$}
	{
		$(N,d) \gets Q.\Extract()$  \Comment*[f]{(node, distance)} \\
		$\Descend(q,N,\True)$  \Comment*[f]{descend again, but check leaves now} \\
		$\ell \gets \ell + 1$ \Comment*[f]{increase leaves checked} \\
	}
	\Return $H$
}

\BlankLine

\Block({$\Descend(q,\Node\ N,\Check)$} \Comment*[f]{descend node $N$ for query $q$})
{
	$d \gets N.\Dist(q)$  \Comment*[f]{signed distance to boundary} \\
	\If(\Comment*[f]{$q$ is in negative half-space}){$d < 0$}
	{
		$Q.\Insert(N.right,|d|)$  \Comment*[f]{remember right child} \\
		$\Descend(q,N.left,\Check)$  \Comment*[f]{descend left child}
	}
	\Else
	{
		$Q.\Insert(N.left,|d|)$  \Comment*[f]{and vice versa} \\
		$\Descend(q,N.right,\Check)$
	}
}

\Block({$\Descend(q,\Leaf\ N,\Check)$} \Comment*[f]{test query $q$ on leaf $N$})
{
	\lIf{$\neg\Check$}{\Return}
	\For{$i \in N.\Points$}
	{
		$H.\Insert(i,\|q-X_i\|^2)$ \Comment*[f]{distances to points $X_i$ in leaf $N$} \\
	}
}
\caption{\kgeraf: searching.\label{alg:search}}
\end{algorithm}

\subsection{Searching}
\label{sec:search}

Searching takes place in parallel in all trees; this does not refer to independent search per tree, but rather that nodes from all trees are visited in a particular order using a shared min-priority queue $Q$. The idea is that given a bound $c$ on the total leaves to be checked, the query iteratively descends the most promising nodes from all trees, and the criterion is the distance of the query to the hyperplane specified by each node.

A shown in Alg.~\ref{alg:search}, the query initially descends all trees of forest $F$ while all visited nodes are stored in $Q$, without checking any leaves. Then, for each node extracted from $Q$, the query descends again, this time computing distances to all points in the leaf. For each decision made at a node while descending, the other one is stored in $Q$. In particular, the \emph{signed} distance $d = N.\textsc{dist}(q)$ of query $q$ to the hyperplane specified by node $N$ is
\begin{equation}
	N.\textsc{dist}(q) = N.tree.f(q)_{N.c} - N.v
\label{eq:sign-dist}
\end{equation}
where $N.tree.f$ is the isometry of the tree where $N$ belongs, and $N.c$, $N.v$ are the split dimension (coordinate) and value of $N$, respectively. One child of $N$ is chosen to descend according to the sign of $d$, and the other is stored in $Q$ with the absolute distance $|d|$ as key. This key is used for priority in $Q$.

Results are stored in a min-heap $H$ that holds up to $k$ points, where $k$ is the number of neighbors to be returned. For each leaf visted, the distance between $q$ and all points stored in the leaf is computed. For each point $X_i$ of the dataset $X$, $H$ is updated dynamically such that it always contains the $k$ nearest neighbors to $q$. The key used for $H$ is the computed (squared) distance $\|q-X_i\|^2$. A separate array keeps track of points encountered so far, such that no distance is computed twice; this detail is not shown in Alg.~\ref{alg:search}.

For each tree built under isometry $f$, the transformed query $f(q)$ is used in all tests at internal nodes, but the initial query $q$ is rather used in all distance computations with points stored at leaves. Similarly, the transformed dataset is used only for building the tree but is not stored. This is possible since the isometry leaves distances unaffected. In practice, unlike~(\ref{eq:sign-dist}), the query is transformed according to isometries of all trees prior to descending.

Although no backtracking occurs, visiting new nodes is an implicit form of backtracking. However, given the bound on the number of leaves to be visited, search is approximate. In particular, apart from the case when $Q$ is empty, search terminates when $c/(1+\epsilon)$ leaves have been checked.
That is, up to $c$ leaves are checked for $\epsilon=0$, while this bound decreases for $\epsilon > 0$, making search faster and less accurate.
\section{Implementation}
\label{sec:impl}

This section discusses our C++ implementation of \kgeraf, which is available online\footnote{\url{https://github.com/gsamaras/kd_GeRaF}}. The project is open source, under the BSD 2-clause license.
Important implementation issues are discussed here, focusing on efficiency.


\head{Parameters.} Our implementation provides several parameters to allow the user to fully customize the data structure and search algorithm:

\begin{itemize}

	\item[$m$] Number of trees in forest. A small number yields fast building and search, but may reduce accuracy; a large $m$ covers space better and enhances accuracy, but slows down building and search.

	\item[$t$] Number of dimensions used for splits. As $d$ increases, a larger $t$ is better, until accuracy begins to drop. The optimum $t$ depends on the input.

	\item[$p$] Maximum number of points per leaf. A large $p$ means short trees, and saves space; a small $p$ accelerates search, but may reduce accuracy.

	\item[$c$] Maximum number of leaves to be checked during search. The higher this number, the higher the accuracy and search time.

	\item[$\epsilon$] Determines search accuracy (Definition~\ref{Danns}); more accurate search comes at the expense of slower query.

	\item[$k$] Number of neighbors to be returned for a query; specified during search.


%

\end{itemize}



\head{Configuration.} We provide a simple and fast automatic configuration method for parameter tuning. Given a dataset and $\epsilon$ we automatically configure all parameters above, except $k$. In particular, taking into account $n, d$ and the five coordinates of greatest variance, we configure parameters $p, c, t, m$, limiting their values to powers of two. The particular values chosen are piecewise constant functions of $\epsilon, n, d$, where constants have been obtained by experience, \ie by manually setting parameters on a number of datasets. This kind of tuning is largely subjective. The runtime is negligible, since the variances are computed by the algorithm anyway. However, the resulting parameter set is not optimal, \eg in terms of accuracy or speed.


\head{Tree structure.} Every tree consists of split nodes and leaves. A split node contains the split dimension and value, while a leaf contains a number of point indices. Points are stored only once, regardless of forest size. We store trees in arrays to benefit from contiguous storage. As discussed in section~\ref{sec:exp}, split value randomization is not beneficial so we disable it. In this case, we split at medial and trees are perfect, thus space is optimized. No re-allocation is needed because we know the size of the tree in advance.

\begin{algorithm}[h]
\SetFuncSty{textsc}
\SetDataSty{emph}
\newcommand{\expr}[1]{$\langle${#1}$\rangle$}
\newcommand{\commentsty}[1]{{\color{DarkGreen}#1}}
\SetCommentSty{commentsty}
\SetKwComment{Comment}{$\triangleright$ }{}
\SetKwBlock{Block}{function}{}
\SetKwInOut{Input}{input}
\SetKwInOut{Output}{output}
\SetKw{Return}{return}
\SetKwFunction{Size}{size}
\Input{sequence $x$ of real vectors in $\real^d$}
\Output{variance on each dimension of the vectors in $x$}
\Begin
{
	\lIf(\Comment*[f]{zero vector in $\real^d$}) {$x.\Size() < 2$}{\Return return $0$}
	$\mu \gets 0$; $v \gets 0$ \Comment*[f]{zero vectors in $\real^d$} \\
	\For{$n \gets 1$ \KwTo $x.\Size()$}
	{
		$\alpha \gets 1/n$ \Comment*[f]{$\alpha$: scalar} \\
		$\delta \gets x[i] - \mu$ \Comment*[f]{$\delta$: vector in $\real^d$} \\
		$\mu \gets \mu + \alpha\delta$ \Comment*[f]{$\mu$: vector in $\real^d$} \\
		$v \gets v + \delta \circ (x[i] - \mu)$ \Comment*[f]{$\circ$: Hadamard product; $v$: vector in $\real^d$} \\
	}
	\Return $v/(n - 1)$
}
\caption{Modified Knuth's online variance algorithm \label{alg:knuth}}
\end{algorithm}

\head{Median, variances.} The median is found efficiently by the quickselect algorithm, with average complexity $O(n)$. Variance is computed by an extension of Knuth's online algorithm~\cite[p.232]{Kn98}, as shown in Alg.~\ref{alg:knuth}. In particular, we extend the algorithm to operate in parallel on a sequence of vectors rather than scalars. In doing so, we replace vector division with scalar $n$ by multiplication with $\alpha = 1/n$. This choice provides significant speed-up.


\head{Distance computation.} This is the most expensive task during search in high dimensions. To speed it up, we note that squared Euclidean distance between point $x$ and query $q$ is $\|q-x\|^2 = \|q\|^2 + \|x\|^2 - 2q^\top x,$ where $\|q\|$ is constant, while $\|x\|$ can be stored for all points. Thus distance computation reduces to dot product, providing a speed-up of $>10\%$ in certain cases. The space overhead is one scalar per point, which is negligible in high dimensions since all points are stored in memory.




\head{Parallelization.} The building process is trivially parallelizable: we just assign building of individual trees to different threads, making sure that the work is balanced among threads. Searching is not performed in parallel: due to use of a single priority queue for all trees, more work would be required for communication between different threads. It would be interesting to investigate this extension in future work.



\section{Experiments}
\label{sec:exp}

\setlength{\tabcolsep}{3pt}
\setlength{\intextsep}{16pt}
\newcommand{\squeeze}{\vspace{-20pt}}
\newcommand{\tg}[1]{{\color{gray}#1}}
\renewcommand{\floatpagefraction}{0.9}

This section presents our experimental results and comparisons on a number of synthetic and real datasets. All experiments are conducted on a processor at 2.40~GHz$\times 4$ with 3.8~GB memory, except for GIST dataset with $n = 10^6$, for which we use a processor at 3~GHz$\times 4$ with 8~GB. We compare to BBD-trees as implemented in {\tt ANN}, LSH as implemented in {\tt E$^2$LSH}, {\tt FLANN}, and our implementation of PQ.

\head{Datasets.} We use five datasets of varying dimensionality and cardinality. To test special topologies, the first two, \emph{Klein bottle} and \emph{Sphere} are synthetic. We generate points on a Klein bottle and a sphere embedded in $\real^d$, then add to each coordinate zero-mean Gaussian noise of standard deviation~0.05 and~0.1 respectively. In both cases, queries are nearly equidistant to all points, which implies high miss rates.

The other three datasets, \emph{MNIST}\footnote{\url{http://yann.lecun.com/exdb/mnist/}}, \emph{SIFT} and \emph{GIST}\footnote{\url{http://corpus-texmex.irisa.fr/}}~\cite{JeDS11}, are common in computer vision and machine learning. MNIST contains vectors of 784 dimensions, that are $28\times 28$ image patches of handwritten digits. There is a set of $60$k vectors, plus an additional set of $10$k vectors that we use as queries. SIFT is a 128-dimensional vector that describes a local image patch by histograms of local gradient orientations. GIST is a 960-dimensional vector that describes globally an entire image. SIFT and GIST datasets each contain one million vectors and an additional set for queries, that are $10^4$ for SIFT and 1000 for GIST. For GIST, we also use the first $10^5$ vectors as a separate smaller dataset.


\head{Parameters.} Most experiments use the default parameters provided by existing implementations but, on specific inputs, we have optimized the parameters manually. This improves performance, but is quite impractical in general. {\tt FLANN} and \geraf determine automatically the parameters given the dataset and $\epsilon$, while {\tt ANN} uses default parameters regardless of $\epsilon$. {\tt E$^2$LSH} provides automatic parameter configuration, but not for the most important one, $R$, used in solving a randomized version of $R$-near neighbor. This is a major drawback, since the user has to manually identify $R$ at every input. As discussed below, accuracy measurements only refer to the first nearest neighbor, so we always set $k=1$ in Alg.~\ref{alg:search}. The same holds for {\tt BBD} and {\tt FLANN}, but not for {\tt LSH} where the number of neighbors is only controlled by $R$.

In \kgeraf, we have observed that rotation does not seem to affect search performance, despite the time penalty, \ie build time increases from 0.26 to 1.35sec on Klein bottle with $n=10^4, d=10^4$. Similarly, split value randomization brings no benefit, despite its cost: build time increases from 0.06 to 1.92 (89.8) sec for approximate (exact) diameter computation, while search accuracy decreases for approximate computation. We have therefore disabled these two randomization factors.

\head{Implementation.} Before presenting experimental comparisons to other methods, we measure the effect of two implementation issues discussed in section~\ref{sec:impl}, in particular parallelization and distance computation. Both are measured on SIFT dataset  with $d=128$. On four cores, parallelization reduces build time from~$15$ to $9$msec for $n=10^4$, and from~$3.32$ to $1.48$sec for $n=10^6$: the speedup is higher for larger forests. On the other hand, reduction of distance computation to dot product reduces build time from~$3.06$ to $2.67\mu$sec per point. However, this approach appears to be effective only when $d > 100$ in practice.


\begin{table}
\begin{center}
\begin{tabular}{|c|cccc|cccc|cccc|} \hline
              & \multicolumn{4}{c|}{Sphere $n = 10^3, d = 10^4$} & \multicolumn{4}{c|}{Klein $n = 10^4, d = 10^2$} & \multicolumn{4}{c|}{MNIST $n = 60$k$, d = 784$} \\ \hline
$\epsilon$    &  0    &  0.1  &  0.5  &  0.9  & 0    & 0.1  & 0.5  & 0.9  &   0   &   0.1 &   0.5 &   0.9  \\ \hline\hline
{\tt BBD}     &  1.25 &  1.26 &  1.30 &  1.25 & 0.13 & 0.14 & 0.17 & 0.14 & 187.5 & 184.3 & 185.1 & 185.6  \\
{\tt LSH}     &  0.21 &  0.16 &  0.18 &  0.31 & 0.11 & 0.07 & 0.03 & 0.05 &  1.47 & 69.76 & 48.47 & 14.35  \\
{\tt FLANN}   & 25.0  & 25.4  & 25.5  & 25.6  & --   & --   & --   & --   & 244.6 & 217.2 & 157.3 & 142.0  \\
\geraf        &  0.06 &  0.06 &  0.06 &  0.06 & 0.06 & 0.06 & 0.06 & 0.08 & 8.167 & 8.567 & 8.579 & 8.565  \\ \hline
\end{tabular}
\caption{Build time (s) for three representative datasets. {\tt FLANN} does not finish after $4$hr on Klein bottle, which is indicated by `--'.}
\label{tab:build}
\end{center}
\squeeze
\end{table}

\head{Preprocessing.} For all methods this includes building, but for {\tt FLANN} and \geraf it also includes automatic parameter configuration. Build time is related to the required precision as expressed by $\epsilon$. For {\tt LSH}, $\epsilon$ is failure probability and its build time is the most sensitive to $\epsilon$. Despite requesting the user to manually determine parameter $R$, {\tt LSH} performs an automatic parameter configuration as well, which is included in the building process.

Table~\ref{tab:build} shows representative experiments. {\tt FLANN} has difficulties with automatic configuration, which does not terminate after $4$hr on Klein bottle and is quite slow in general. {\tt LSH} is unexpectedly fast on MNIST for $\epsilon=0$, which may be due to the parameters chosen by auto-tuning. \geraf works well with automatic configuration, and is typically one order of magnitude faster than other methods. Its preprocessing time may increase with $\epsilon$ since this requires fewer points per leaf, hence more subdivisions.

We additionally carry out an experiment with product quantization (in particular, IVFADC)~\cite{JeDS11} on SIFT, implemented on Matlab with {\tt Yael}\footnote{\url{https://gforge.inria.fr/projects/yael}} library. Its off-line processing includes codebook learning, which takes 440sec for 50 $k$-means iterations and encoding/indexing, which takes 45sec. The latter time is competitive if codebooks are existing from a similar dataset, but the total time given a new unknown dataset is quite higher than \geraf and {\tt LSH}; and even higher than {\tt FLANN} with default values.

\begin{table}
\begin{center}
\scriptsize
\begin{tabular}{|c|rrrr|rrrr|rrrr|} \hline
              & \multicolumn{4}{c|}{Sphere $n = 10^3, d = 10^4$} & \multicolumn{4}{c|}{Klein $n = 10^4, d = 10^2$} & \multicolumn{4}{c|}{Sphere $n = 10^5, d = 10^2$} \\ \hline
$\epsilon$    &  0     &  0.1   &  0.5       &  0.9   &  0     &  0.1   &  0.5   &  0.9   &  0     &  0.1   &  0.5   &  0.9    \\ \hline\hline
              & \multicolumn{12}{c|}{miss \%} \\ \hline
{\tt BBD}     &  0     &  100   &  100   &  100   &  0     &  59    &  59    &  59    &  1     &  100   &  100   &  100    \\
{\tt LSH}     &  45    &  45    &  45    &  45    &  1     &  1     &  20    &  63    &  2     &  2     &  2     &  2      \\
{\tt FLANN}   &  0     &  0     &  0     &  0     &  --    &  --    &  --    &  --    &  100   &  100   &  100   &  100    \\
\geraf        &  0     &  24    &  24    &  100   &  2     &  3     &  3     &  5     &  2     &  26    &  40    &  81     \\ \hline\hline
& \multicolumn{12}{c|}{search (ms)} \\ \hline
{\tt BBD}     &  9.100 & \tg{0.210} & \tg{0.220} & \tg{0.200} &  0.470 &  0.043 &  0.046 &  0.052 &    12      & \tg{0.024} & \tg{0.028} & \tg{0.026} \\
{\tt LSH}     & 17.000 &    16.000  &    18.000  &    17.000  &  2.700 &  2.400 &  1.900 &  0.850 &    28.000  &    24.000  &    22.000  &    22.000  \\
{\tt FLANN}   &  0.310 &     0.280  &     0.350  &     0.320  &  --    &  --    &  --    &  --    & \tg{0.021} & \tg{0.021} & \tg{0.020} & \tg{0.021} \\
\geraf        &  0.400 &     0.200  &     0.150  & \tg{0.100} &  0.100 &  0.083 &  0.083 &  0.070 &     3.900  &     2.900  &     1.500  &     1.300  \\ \hline
\end{tabular}
\caption{Search accuracy and times for synthetic datasets. Search times in gray represent failure cases where miss rate is 100\%. Queries are nearly equidistant to points, which explains high miss rates, especially for {\tt BBD} and {\tt FLANN}; `--' indicates preprocessing does not finish after $4$hr.}
\label{tab:search}
\end{center}
\squeeze
\end{table}


\head{Search.} We report query times and miss rates for four representative values of $\epsilon$. The miss rate is the percentage of queries where the reported neighbor is \emph{not} the exact one. In case of ties, any point at the same distance as the nearest neighbor is accepted as correct. Table~\ref{tab:search} shows results for all methods on three representative synthetic datasets. {\tt BBD} and {\tt FLANN} have problems with high miss rate or having failed in automatic preprocessing. {\tt LSH} is at least one order of magnitude slower than \geraf. In most cases \geraf is faster (especially for large $d=10^4$), with competitive miss rate, except for {\tt FLANN} on Sphere with $d=10^4$, which is the best dimension for {\tt FLANN}.

\setlength{\tabcolsep}{0pt}
\pgfplotsset{
	every axis x label/.style={at={(ticklabel cs:0.5)},anchor=north},
	every axis y label/.style={at={(ticklabel cs:0.5)},rotate=90,anchor=south},
}

\begin{figure}
\begin{tabular}{cc}
\begin{tikzpicture}
	\begin{semilogyaxis}[
		font = \scriptsize,
		width = .5\linewidth,
		title = {(a) MNIST $n = 6\times 10^4, d = 784$},
		xlabel = {miss \%},
		ylabel = {search (sec)},
		legend entries = {{\tt LSH}, {\tt FLANN}, \geraf},
		label/.style={nodes near coords,point meta=explicit symbolic},
		ori/.style={every node near coord/.append style={anchor=#1}},
		ori=south west,
	]
	\pgfplotstableread{
		miss  t_secs   epsilon  build
		0     0.066985 0        1.474963
		0     0.039804 0.5      48.477440
		3.8   0.017461 0.9      14.353137
	}{\lsh}
	\pgfplotstableread{
		miss  t_secs      epsilon  build
		20.8  0.000064915 0        244.679469064
		23.9  0.000054398 0.1      217.227075346
		29.6  0.000044363 0.5      157.398897331
		56.6  0.000032418 0.9      142.069667287
	}{\flann}
	\pgfplotstableread{
		miss  t_secs     epsilon  build
		 4.4  0.00424701 0        8.16765
		10.3  0.0023728  0.1      8.56714
		14.2  0.0022603  0.9      8.56524
	}{\rkdf}
	\pgfplotstableread{
		miss  t_secs   epsilon  build
		0     0.053233 0        187.595490
		0     0.048893 0.1      184.375523
		0     0.033881 0.5      185.193572
		6     0.023884 0.9      185.697547
	}{\bbd}
	\addplot[label,mark=*,blue] table[meta=epsilon] \lsh;
	\addplot[label,mark=square*,red] table[meta=epsilon] \flann;
	\addplot[label,ori=north,mark=triangle*,magenta] table[meta=epsilon] \rkdf;
	\end{semilogyaxis}
\end{tikzpicture}
&
\begin{tikzpicture}
	\begin{semilogyaxis}[
		font = \scriptsize,
		width = .5\linewidth,
		title = {(b) SIFT $n = 10^6, d = 128$},
		xlabel = {miss \%},
		xmax = 85,
		legend entries = {{\tt LSH}, {\tt FLANN}, \geraf},
		label/.style={nodes near coords,point meta=explicit symbolic},
		ori/.style={every node near coord/.append style={anchor=#1}},
		ori=south west,
	]
	\pgfplotstableread{
		miss t_secs epsilon
		nan  nan    0
		nan  nan    0.1
		1.34 0.12   0.5
		8.61 0.035  0.9
	}{\lsh}
	\pgfplotstableread{
		miss  t_secs      epsilon
		71.23 0.000058147 0-0.9
	}{\flann}
	\pgfplotstableread{
		miss t_secs epsilon anchor
		1  0.019 0   north
		4  0.012 0.1 north
		8  0.009 0.5 north
		22 0.003 0.9 south
	}{\rkdf}
	\pgfplotstableread{
		miss  t_ms par
		14.61 1.83 4/100
		 7.73 3.52 8/100
		 4.32 4.31 8/1000
		 1.01 7.56 16/1000
	}{\pq}

	\addplot[label,mark=*,blue] table[meta=epsilon] \lsh;
	\addplot[label,ori=south,mark=square*,red] table[meta=epsilon] \flann;
	\addplot[label,ori=north east,mark=triangle*,magenta] table[meta=epsilon] \rkdf;
	\end{semilogyaxis}
\end{tikzpicture}
\\
\begin{tikzpicture}
	\begin{semilogyaxis}[
		font = \scriptsize,
		width = .5\linewidth,
		title = {(c) GIST $n=10^5, d=960$},
		xlabel = {miss \%},
		ylabel = {search (sec)},
		ymin = .01, ymax = .3,
		legend entries = {{\tt LSH}, \geraf},
		label/.style={nodes near coords,point meta=explicit symbolic},
		ori/.style={every node near coord/.append style={anchor=#1}},
		ori=south west,
	]
	\pgfplotstableread{
		miss t_secs epsilon anchor
		0.5 0.24  0  east
		0.5 0.21 0.1 west
		0.5 0.18 0.5 east
	}{\lsh}
	\pgfplotstableread{
		miss t_secs epsilon
		1 0.056  0
		1 0.025 0.1
		2 0.024 0.5
		3 0.024 0.9
	}{\rkdf}
	\pgfplotstableread{
		miss  t_ms    par
		38.1   5.22   4/100
		23.2  10.04   8/100
		18.5  13.26   8/1k
		 6.6  20.45  16/1k
		 6.6  43.69  16/10k
		 1.4  68.14  32/10k
	}{\pq}
	\addplot[label,ori=west,mark=*,blue] table[meta=epsilon] \lsh;
	\addplot[label,ori=south,mark=triangle*,magenta] table[meta=epsilon] \rkdf;
	\end{semilogyaxis}
\end{tikzpicture}
&
\begin{tikzpicture}
	\begin{semilogyaxis}[
		font = \scriptsize,
		width = .5\linewidth,
		title = {(d) GIST $n=10^6, d=960$},
		xlabel = {miss \%},
		legend entries = {{\tt LSH}, {\tt FLANN}, \geraf},
		legend pos = {south east},
		label/.style={nodes near coords,point meta=explicit symbolic},
		ori/.style={every node near coord/.append style={anchor=#1}},
		ori=south west,
	]
	\pgfplotstableread{
		miss t_secs epsilon anchor
		100 1.3109  0  east
 		100 1.065155 0.1-0.9 east
	}{\lsh}
	\pgfplotstableread{
		miss t_secs      epsilon
		40.7 0.028099477 0
		42.2 0.005465471 0.1-0.9
	}{\flann}
	\pgfplotstableread{
		miss t_secs epsilon
 		3.6  0.25   0
		11.5 0.11   0.1
		22.5 0.06   0.9
	}{\rkdf}
	\pgfplotstableread{
		m        t_ms    par
		42.60   5.60    4/100
		36.70  10.49    8/100
		17.30  15.00    8/1k
		10.10  23.03   16/1k
		 5.00  53.10   16/10k
		 0.80  73.01   32/10k
	}{\pq}
	\addplot[label,ori=east,mark=*,blue] table[meta=epsilon] \lsh;
	\addplot[label,mark=square*,red] table[meta=epsilon] \flann;
	\addplot[label,ori=west,mark=triangle*,magenta] table[meta=epsilon] \rkdf;
	\end{semilogyaxis}
\end{tikzpicture}
\end{tabular}
\caption{Search accuracy and times on real datasets. In (b), {\tt LSH} is out of memory for $\epsilon \in \{0, 0.1\}$. In all cases, {\tt BBD} is out of memory and {\tt FLANN} does not preprocess after $4$hr for any $\epsilon$. Its measurements in (a),(b),(d) refer to manually configured parameters.}
\label{fig:siftgist}
\end{figure}
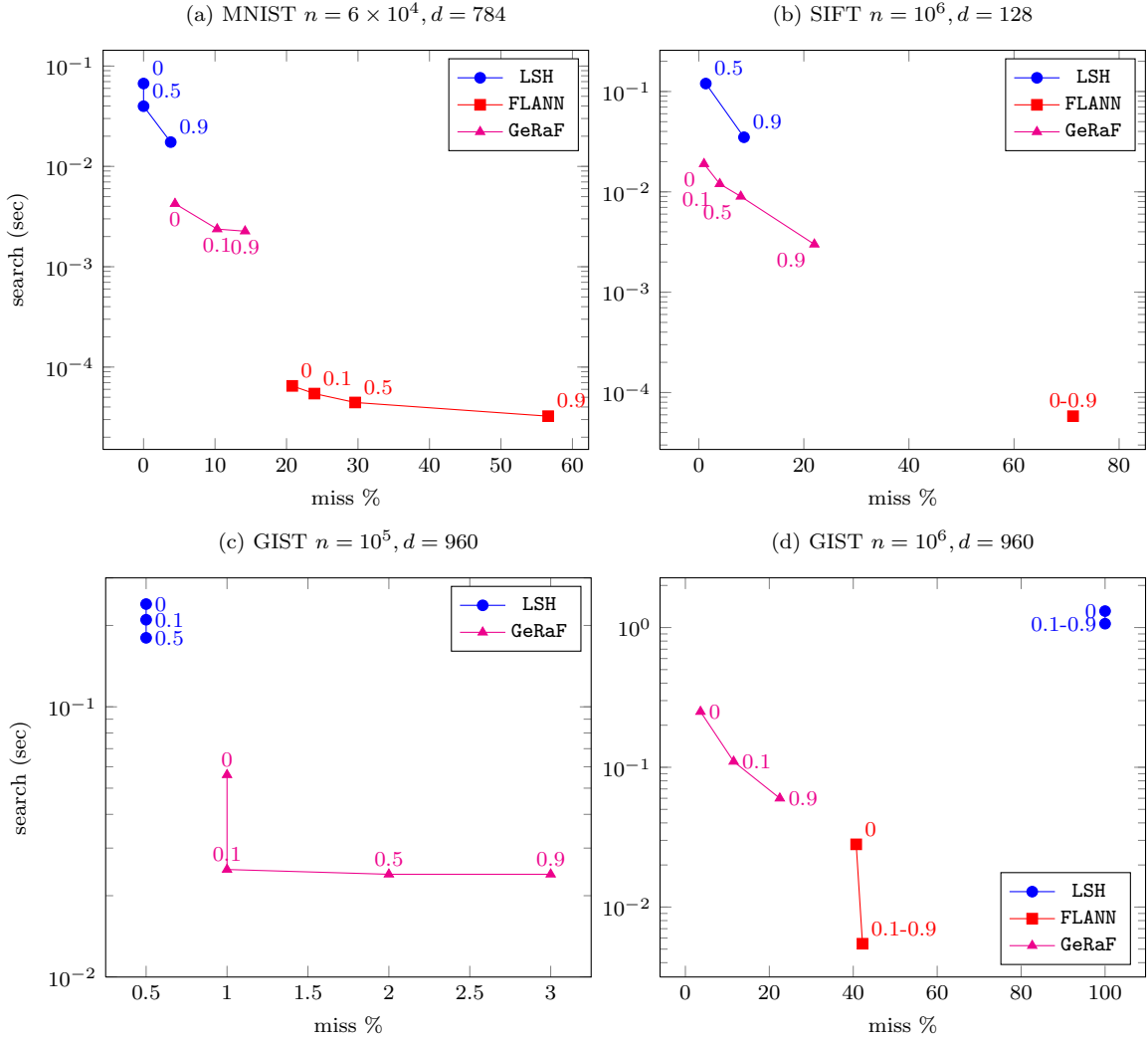

Figure~\ref{fig:siftgist} presents four representative datasets with real data. {\tt BBD} and {\tt FLANN} have problems with either running out of memory or not completing automatic build. \geraf is typically faster than {\tt LSH} by at least an order of magnitude at the same accuracy. In all cases, {\tt FLANN} preprocessing does not terminate after $4$hr so we manually configure parameters because default ones yield even higher miss rates. {\tt FLANN} is generally the fastest method but with low accuracy. On GIST, with $d=960$, \geraf shows best performance. {\tt LSH} has $0.5\%$ better miss rate for $n=10^5$, but is quite slower; it also fails with $100\%$ miss rate for $n=10^6$. With automatic configuration, \geraf always yields a good trade-off between accuracy and speed.

We also experiment with PQ (IVFADC) in these datasets, which is known to outperform {\tt FLANN}~\cite{JeDS11} when combined with re-ranking. For instance, it takes $7$ ($70$) msec for a miss rate of $1$\% on SIFT (GIST $n=10^6$). However its training is slow, as noted above.

\setlength{\tabcolsep}{3pt}

\begin{table} \begin{center}
\begin{tabular}{ |c|c||rrrr|rrrr| } \hline
	\multirow{2}{1em}{$n$} & \multirow{2}{1em}{$d$} & \multicolumn{4}{c|}{miss \%} & \multicolumn{4}{c|}{search ($\mu$sec)} \\ \cline{3-10}
	 & & {\tt BBD} & {\tt LSH} & {\tt FLANN} & \geraf & {\tt BBD} & {\tt LSH} & {\tt FLANN} & \geraf\\ \hline\hline
	\multirow{4}{2em}{$10^3$} & 100    & 100 & 0  & 16   & 0  & \tg{1}     & 212   & 12         & 199   \\
	                          & 1000   & 100 & 50 & 100  & 50 & \tg{5}     & 1850  & \tg{34}    & 14    \\
	                          & 5000   & 100 & 0  & 100  & 0  & \tg{39}    & 8675  & \tg{149}   & 122   \\
	                          & 10000  & 100 & 37 & 100  & 2  & \tg{276}   & 17000 & \tg{289}   & 520   \\ \hline
	1000   & \multirow{3}{2em}{$10^3$} & 100 & 50 & 100  & 50 & \tg{5}     & 1850  & \tg{34}    & 14    \\
	10000  &                           & 100 & 0  & 100  & 0  & \tg{5}     & 1780  & --         & 390   \\
	100000 &                           & 100 & 8  & 100  & 0  & \tg{276}   & --    & --         & 10900 \\ \hline
\end{tabular}
\caption{Klein bottle search for $\epsilon = 0.1$, for varying $n$ or $d$, where the other parameter is fixed. Search times in gray represent failure cases where miss rate is 100\%. Queries are nearly equidistant from the points, which explains high miss rates. `--' indicates preprocessing does not finish after $2$hr.\label{fixn}}
\end{center}
\squeeze
\end{table}

Table~\ref{fixn} displays, for all methods, the miss rate and search time as a function of $n$ or $d$ when the other parameter is fixed. In cases where miss rate is not $100$\%, \geraf is an order of magnitude faster. The only exception is $d=100$, where the situation is inversed with {\tt FLANN}.



\head{Approximate search evaluation.} We also measure for \geraf the percentage of queries where the reported nearest neighbor does \emph{not} lie within $1+\epsilon$ of the nearest distance. This a more natural measure than miss rate when approximate search is requested given a specific $\epsilon$. For Klein bottle with $n = 10^4, d = 10^2$, this rate is $2\%$ and $0\%$, for $\epsilon = 0$, and $\epsilon\in\{ 0.1, 0.5, 0.9\}$, respectively. In order for the output to always lie within $1+\epsilon$ of optimal, one may set $c = n$, thus disabling the termination condition of leaves to be checked. However, due to the curse of dimensionality, performance nearly reduces to brute force in this case. For GIST with $n = 10^5, d = 960$ for instance, search takes 140ms, whereas miss rate is $0\%$
and $0.4\%$ for $\epsilon = 0$ and $\epsilon\in\{0.1, 0.5, 0.9\}$ respectively.


\begin{table}
\begin{center}
\begin{tabular}{|c|rrrrrrrr| } \hline
		$p$         & 256    & 128    & 64      & 32      & 16      &  4      &  2      &  1      \\ \hline\hline
		build (s)   & 0.0592 & 0.0618 &  0.0674 &  0.0695 &  0.0860 &  0.1159 &  0.1543 &  0.1587 \\
		search (ms) & 0.2324 & 0.1863 &  0.1198 &  0.0941 &  0.0712 &  0.0592 &  0.0743 &  0.0928 \\
		miss \%     & 1      & 1      &  2      &  7      &  6      & 10      & 14      & 22      \\ \hline
\end{tabular}
\caption{\geraf build and search measurements for Klein bottle dataset with $n = 10^4, d = 10^2$ for varying points per leaf $p$.}
\end{center}
\label{tab:ppl}
\squeeze
\end{table}

\head{Points per leaf.} Finally, we measure the effect of storing multiple points per leaf on the Klein bottle dataset. The results are shown in Table~\ref{tab:ppl}. It is clear that search time improves when there are less points per leaf, and this is why a single point per leaf is a common approach. However, the build time and most importantly the miss rate also increase significantly. We therefore provide a reasonable trade-off by automatically adjusting parameter $p$.
\section{Discussion}
\label{sec:discuss}

We have presented an efficient data structure for approximate nearest neighbor search that explores different randomization strategies, and an efficient implementation, \geraf, that is found competitive against existing implementations of several state-of-the-art methods. We provide a simple but effective automatic parameter configuration that yields the fastest preprocessing, including both configuration and building, as well as a successful trade-off between accuracy and speed. Most competing methods have difficulties in running out of memory at large scale (\eg, {\tt BBD}), slow or non-terminating parameter configuration (\eg, {\tt FLANN}), or unstable search behavior between accurate (but slow) or fast (but inaccurate) search (\eg, {\tt LSH} and {\tt FLANN}). PQ is consistently faster and more accurate at search, but is significantly slower to build, which is impractical when the dataset is updated. Our findings are consistent on both synthetic and real datasets of a wide range of dimensions and cardinalities.


Interesting open questions include whether and how \geraf can be fully dynamic, supporting insertions and deletions, as well as handling batch queries in an optimized manner. Other future directions include performing parallel or distributed search and more principled parameter configuration with discrete optimization. In fact, recent experiments with parameter tuning by generic algorithms indicate that build time for large datasets such as SIFT can drop by a factor of 100 without significantly affecting search time while reducing miss rate~\cite{Gia15}.



\bibliographystyle{plain}
\bibliography{tex/aem}

\end{document}